\documentstyle[aps,epsfig,manuscript]{revtex}

\begin{document}

\title{Transition to plastic motion as a critical phenomenon and
anomalous interface layer of a 2D driven vortex lattice}
\author{L. Fruchter}
\address{Laboratoire de Physique des Solides,
Universit\'{e} Paris-Sud, C.N.R.S., B\^{a}t. 510, 91405 Orsay
France}

\date{\today}
\maketitle

\begin{abstract}
The dynamic transition between the ordered flow and the plastic
flow is studied for a two-dimensional driven vortex lattice, in
the presence of sharp and dense pinning centers, from numerical
simulations. For this system, which does not show smectic
ordering, the lattice exhibits a first order transition from a
crystal to a liquid, shortly followed by the dynamical transition
to the plastic flow. The resistivity provides a critical order
parameter for the latter, and critical exponents are determined
in analogy with a percolation transition. At the boundary between
a pinned region and an unpinned one, an anomalous layer is
observed, where the vortices are more strongly pinned than in the
bulk.
\end{abstract}


\narrowtext

\section{Introduction}

Following extensive studies on the effect of disorder on the
static vortex lattice, the physics of the vortex lattice with
random quenched disorder and driven by a uniform force has
attracted recently much attention. Interacting systems, forming
periodic structures at the equilibrium, were already the subject
of much interest since the earlier studies of charge density
waves\cite{fukuyama72,fisher85}. The complexity of the depinning
phenomena was soon pointed out, in the sense that the description
of the depinning threshold by a critical phenomena is no longer
valid when one takes into account the possibility of topological
defects within the periodic
structure\cite{fisher85,coppersmith91}. Plasticity, which is
commonly observed closed to the depinning threshold, is a dramatic
illustration in the case of the two dimensional vortex lattice.
There has been several investigations of the driven vortex 'phase
diagram' which have enriched the canonical description
\cite{koshelev94}: \textit{pinned vortex glass - plastic flow -
moving crystal} as the driving force is increased. Amongst these,
numerical simulations of two dimensional vortex assemblies,
initiated by the work of Brandt\cite{brandt83}, have very often
accompanied theoretical progress on the subject. After the
proliferation of the lattice defects was put into evidence,
suggesting a dynamic first order melting transition at the
occurrence of plastic flow\onlinecite{koshelev94,yaron95},
numerical simulations identified the ordered phase as a moving
transverse Bragg glass\cite{moon96}, in agreement with theoretical
expectations\cite{ledoussal98}. Simultaneously, Ryu et al showed
that an hexatic order parameter exhibits a sharp transition as
one enters the moving glass\cite{ryu96}. Later, Olson et al
showed that, for a soft flux lattice ($A_{V} \leq 1$), the
ordered phase presents a smectic order\cite{olson98}. Finally,
Kolton et al recently introduced a 'frozen transverse solid'
beyond the smectic regime, characterized by a drop of the Hall
noise\cite{kolton99}. In a general way, there is often some
confusion about the exact nature of the 'transition'. 'Dynamical
transition' and 'phase' are often employed in place of
'crossover' or 'dynamic regime', without further justification.
Indeed, there seems to be up to now only one strong suggestion of
a genuine dynamical transition in the works in refs
\onlinecite{koshelev94} and \onlinecite{ryu96}. The fact that the
notion of dynamical transition is itself defined only with
difficulty (see ref. \onlinecite{fisher85}) has certainly
contributed to this situation. It is not clear, for instance,
wether one should try to use some dynamical quantity - such as the
correlation length of the local velocities - as was done in ref.
\onlinecite{fisher85}, or if one should use some instantaneous,
topological one - such as the concentration of defects or the
hexatic order parameter in refs \onlinecite{koshelev94} and
\onlinecite{ryu96} - in the search for an order parameter. Here,
it is shown that a simple system, not showing any intermediate
smectic order between the ordered and the plastic flow, exhibits
a second order like transition to the plastic regime.

\section{Experimental details}
\label{experiment}

A two dimensional lattice subjected to a uniform driving force
(applied along the $y$-axis, thereafter denoted longitudinal
direction) in the presence of pins is simulated, using the force
equation :
\begin{equation}
\mathbf{f_{vv}}(r) + \mathbf{f_{p}}(r) + \mathbf{f_{B_{0}}}(x) +
\mathbf{J}\wedge\mathbf{\Phi_{0}} - \eta \;\mathbf{\dot{r}} = 0
\end{equation}
The geometry is analogous to the one of a Corbino disk
experiment: the two edges at $y=const.$ are submitted to a
periodic boundary condition; the ones at $x=const.$ are submitted
to an external magnetic field, $B_{0}$, which is simulated by an
extra force $f_{B_{0}}$ acting on each vortex, perpendicular to
the edges. The force,$f_{B_{0}}(x)$, acting on a vortex at a
distance $x$ from the edge, is that imposed by a semi-infinite
vortex lattice at a distance $a_{0}+x$, where
$a_{0}=\left(\Phi_{0}/B\right)^{1/2}$ is the flux lattice spacing
at the equilibrium.  Flux lines are assumed rigid rods and the
force per unit length between vortices separated by a distance
$r$ is\cite{brandt83a}:
\begin{equation}
f_{vv}(r)=\left(A_{V}/\lambda\right) K_{1}\left(r/\lambda\right)
\label{fvv}
\end{equation}
where $K_{1}$ is a Bessel function. This is strictly a good
approximation only in the case of vortex lines (rods) and for 2D
vortices a logarithmic interaction should be used. The
interaction between vortices was cut at a distance $5\lambda$,
using an interpolation to zero. This was done in order to avoid
spurious distortions of the equilibrium lattice from the
Abrikosov lattice or the introduction of topological defects, as
was shown to occur for a sharp cutoff in ref.
\onlinecite{fangohr00}.

The sample dimensions were $100\:a_{0}$ along $x$-axis and
$70\:a_{0}$ along $y$-axis. Strong pinning centers are randomly
distributed in the sample. A pin free region was left for $x <
25\:a_{0}$ and $x > 75\:a_{0}$. Doing so, a defect free lattice
is obtained at the edges of the sample, providing well defined
boundary conditions. The density of the pinning sites is
$n_{P}=B_{\Phi}/\Phi_{0}$, with $\Phi_{0}$ the flux quantum and
$B_{\Phi}$ the 'matching field' for which an equilibrium flux
line lattice shows the density of flux lines $n_{V}$. The force
per unit length exerted by a pin at a distance $r$ from the line
is given by:
\begin{equation}
f_{p}(r) = \left(2\;A_{P}/r_{P}\right)\:(r/r_{P});\: \textrm{for
$r \leq r_{P}$},\;
 0  \;\textrm{for $r > r_{P}$}
\label{fp}
\end{equation}
The pinning force is exactly balanced by the Lorentz force,
$\mathbf{J}\wedge\mathbf{\Phi_{0}}$, for $J = J_{0} =
2\;A_{V}/r_{P}\;\Phi_{0}$ (in the following, $j=J/J_{0}$). In the
present study, the following parameters were used : $\lambda =
1.57\;a_{0}$, $r_{P}=4.9\;10^{-2}\;a_{0}$,
$A_{P}/A_{V}=2.5\;10^{-2}$ and $B_{\Phi}=6\;B$. Using the
notations in \onlinecite{brandt83}, this corresponds to sharp,
dense and strong ($A_{P}/r_{0}\;a_{0}\;c_{66} \gg 1$) pins. The
sample contained approximately $N_{V}=7 000$ vortices and 25 000
pins.

\section{Results and Discussion}

\begin{figure}
\begin{center}
\epsfig{file=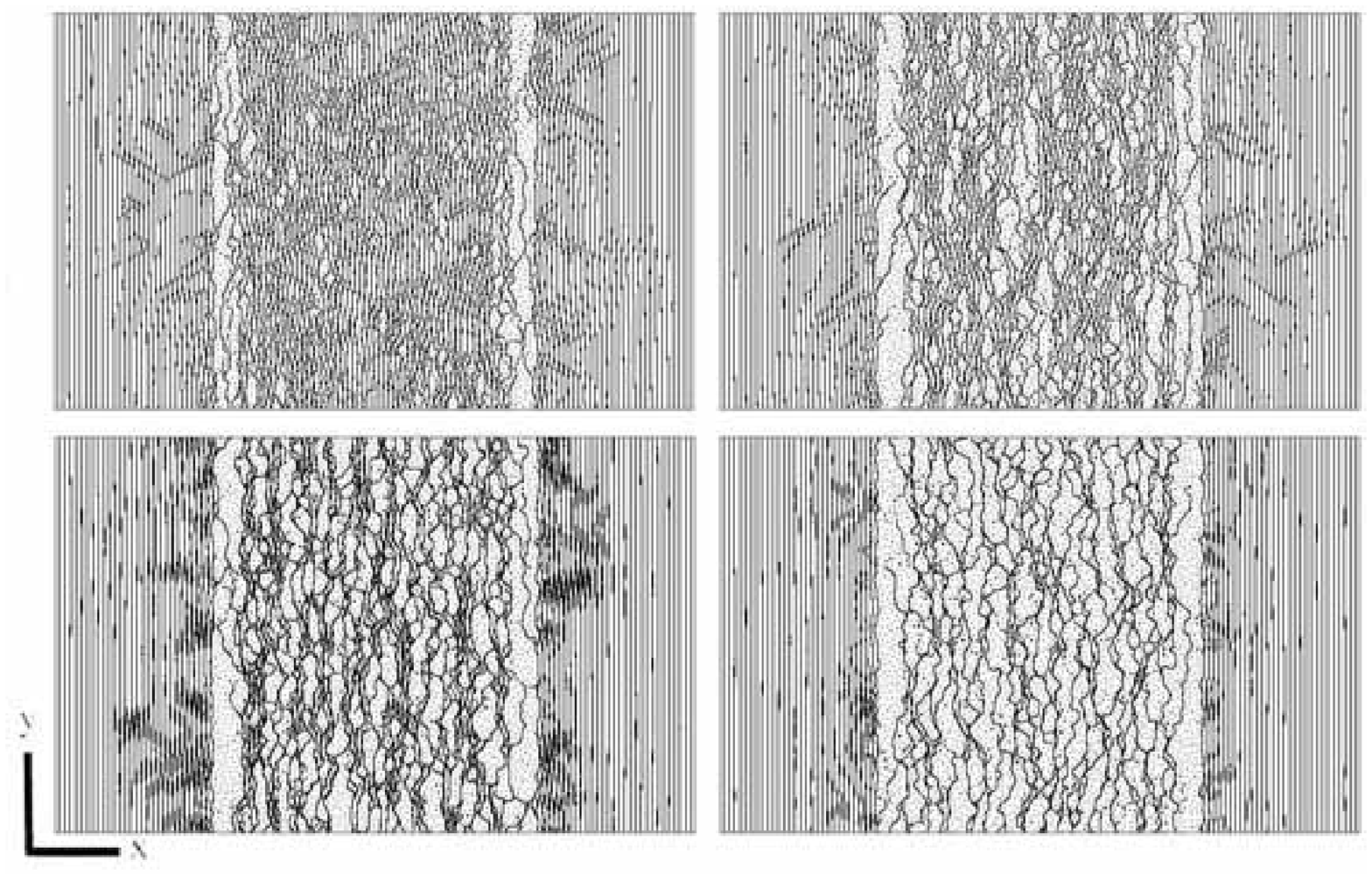, width=10cm}
\end{center}
\caption{Vortices trajectories under uniform driving current
density applied along $x$-axis. Top left and right : $j = 0.31$
and $0.295$, bottom : $0.27$ and $0.24$} \label{phases}
\end{figure}

\begin{figure}
\begin{center}
\epsfig{file=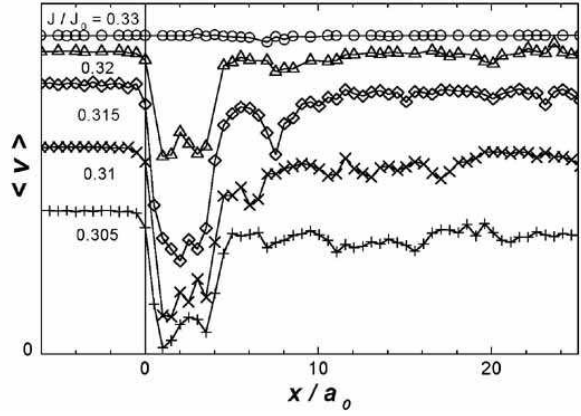, width=8.4cm}
\end{center}
\caption{Average $y$-velocity component profile. The pinning
centers density is non zero where $x \geq 0$. For clarity,
results for each curve were rescaled along the vertical axis.}
\label{vitesse}
\end{figure}

\begin{figure}
\begin{center}
\epsfig{file=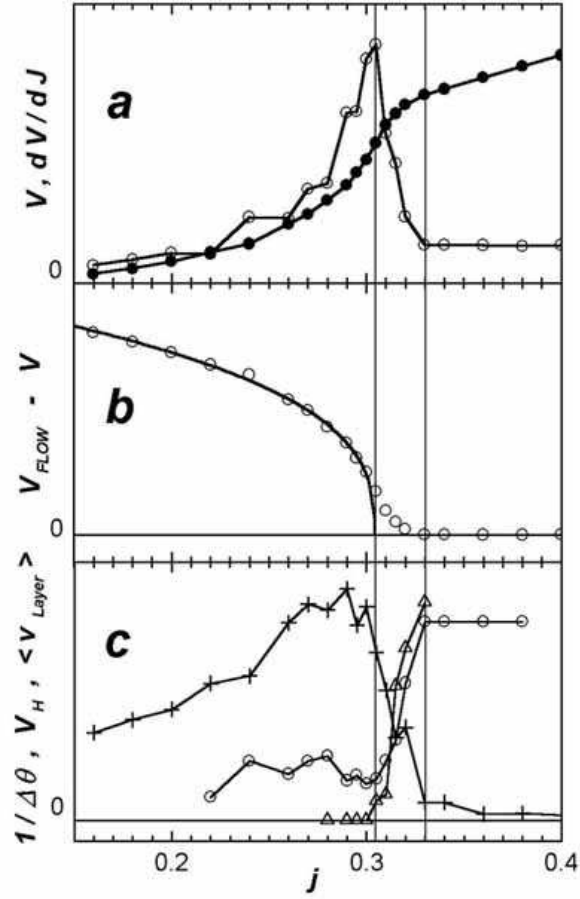, width=8.4cm}
\end{center}
\caption{ \textit{\textbf{a}}, filled : average voltage along the
main flow, computed in the pinned region, away from the anomalous
layer; \textit{\textbf{a}}, empty : voltage derivative.
\textit{\textbf{b}}, points: difference between voltage in
\textit{\textbf{a}} and the free flux flow voltage;
\textit{\textbf{b}}, line : fit to $V_{0} (1-j/j_{C})^{\beta}$,
where $j_{C} = 0.304 \pm 0.001$ and $\beta=0.34 \pm 0.02$.
\textit{\textbf{c}}, crosses: transverse Hall noise;
\textit{\textbf{c}}, circles: average velocity in the anomalous
layer, normalized to that of the bulk;
\textit{\textbf{c}},triangles: inverse of the width of the
diffraction peak at $k=(a_{0},0)$. The vertical line at $j=0.304$
marks the critical driving current, as given by the fit in
\textit{\textbf{b}}; the one at $j=0.33$, the onset of departure
from the free flux flow potential in \textit{\textbf{a}}.}
\label{rau}
\end{figure}

\begin{figure}
\begin{center}
\epsfig{file=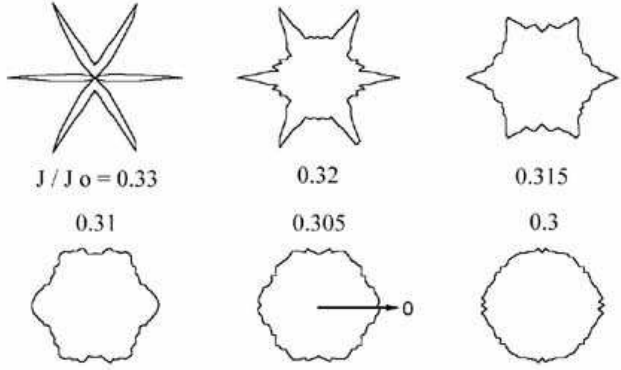, width=8.4cm}
\end{center}
\caption{Polar plot (logarithmic units) of the instantaneous
structure factor, after radial integration over an annulus which
overlaps the diffraction peaks in the first Brillouin zone (the
flux lattice is sampled in the pinned region, away from the
anomalous layer). The zero angle axis points along the reciprocal
direction transverse to the average flux flow.} \label{polaire}
\end{figure}

 Vortices trajectories are shown in Fig.\ref{phases}.
At first sight, they display a striking feature: as the driving
current decreases and the trajectories evolve from correlated
channels to branched trajectories, the vortices are first pinned
at the interfaces between the pinned and the unpinned region.
This is in contradiction with the intuition gained from fluid
dynamics physics, where one would expect the average velocity of
the fluid to decrease monotonously from the one for unpinned
vortices to the one of vortices slowed down by solid friction.
Rather, as shown in Fig.\ref{vitesse}, the average velocity first
drops to a minimum right at the interface between the pinned and
the unpinned region, and then grows to some roughly uniform value
at a distance $\approx 5\:a_{0}$ from the interface. The magnitude
of this anomalous boundary layer effect may be measured as the
ratio of the average velocity in the layer, to the one far away
in the pinned region (Fig.\ref{rau}\textit{\textbf{c}}). Dynamics
regimes were characterized using physical quantities as commonly
done in flux lattice simulations
\onlinecite{moon96,olson98,kolton99}. As shown in Fig.\ref{rau},
the system exhibits a sharp departure from a linear $V-I$
characteristic; an onset of the voltage noise measured in the
direction transverse to the average flux flow; an onset of the
lattice diffraction peaks widening as well as the onset of the
anomalous layer effect at $j_{1}\simeq0.33$. Close to this value,
at $j_{2}\simeq 0.305$, the voltage derivative, $dV/dJ$, shows a
sharp peak; diffraction peaks vanish and the layer effect
saturates. The analysis of the structure factor $S(\mathbf{k})=
n_{V}^{-1}\;|\sum_{i}\:e^{i\;\mathbf{k}\;\mathbf{r_{i}}}|^{2}$ on
an annulus which overlaps the first Brillouin zone diffraction
peaks (Fig.\ref{polaire}) shows the progressive evolution of the
central region of the sample from a well ordered hexagonal
lattice at $j=j_{1}$ to a liquid at $j_{2}$. In between, there is
no evidence in the diffraction intensity for an asymmetry between
the average flux flow direction and the one transverse to it.
Following ref.\cite{olson98}, the regime at $j\leq j_{2}$ is that
of the plastic flow of the amorphous solid. The transition region
$j_{2}<j<j_{1}$ between the plastic regime and the ordered state
differs from the ones described in \cite{olson98} or
\cite{kolton99}, as we find no evidence for the asymmetry needed
in the diffraction intensity for a smectic order or an order
intermediate between a smectic and a crystal. Also, the
transition regime width observed here is only about $10 \%$ of
the critical value for the plastic to quasi-ordered regime
current, while values larger than $30\%$ were found in
\cite{olson98}. These differences are due to parameters much
different from the ones used in \cite{olson98}. Here, pinning
sites are dense and almost point like ($n_{P}/n_{V} = 6$ and
$ab/r_{P} = 25$), while the pinning density is comparable to the
vortex density and pinning sites are extended in
\cite{olson98}($n_{P}/n_{V} = 1.4$ and $ab/r_{P} = 4$). As a
result, vortices do not sense here the asymmetry of the pinning
potential (when it is tilted by the driving force) as they do for
extended defects, and the smectic regime does not occur.

\begin{figure}
\begin{center}
\epsfig{file=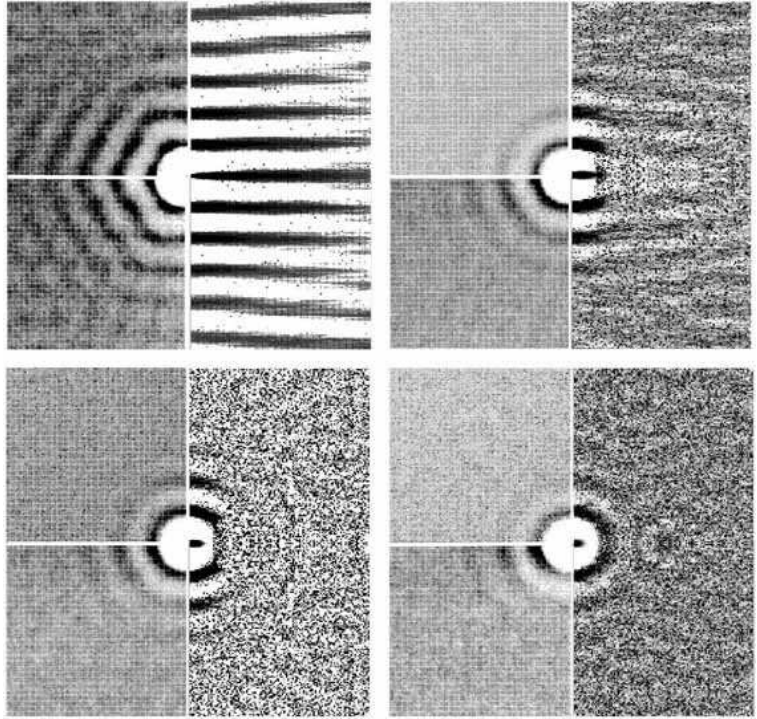, width=10cm}
\end{center}
\caption{Gray scale maps of autocorrelation functions. Clockwise:
$j=0.31,\;0.295,\;0.27,\;0.24$. Left upper quadrant:
autocorrelation function of the instantaneous lattice,
$\mathcal{C}(\mathbf{k})$. Left lower quadrant: autocorrelation
function of the instantaneous lattice, weighted by the
$y$-velocity component, $\mathcal{C}_{V}(\mathbf{k})$. Right half
: autocorrelation function of the vortices trajectories, weighted
by the $y$-velocity component, $\mathcal{C}_{C}(\mathbf{k})$ . The
average flux flow is horizontal.} \label{autocor}
\end{figure}

Within the plastic regime, the evolution of the channels
resembles that of a percolation transition, and the transition
between the ordered flow and the plastic regime may be viewed as
the percolation of dynamic flux channels in the transverse
direction. This similarity was already noticed earlier in
\cite{ryu96}. The analysis of the resistivity quantitatively
demonstrates the validity of a critical phenomenon approach. As
seen in Fig.\ref{rau}\textit{\textbf{b}}, the resistivity may be
fitted to a critical order parameter of the form
$(1-j/j_{C})^{\beta}$, with $\beta = 0.34 \pm 0.02$. The critical
driving force obtained in this way, $j_{C}$, is within fitting
uncertainty identical to $j_{2}$. The restricted intermediate
regime, as observed here, might be crucial for the observation of
the critical behavior, as it tends to smear out the transition.
The interpretation of the anomalous layer effect - which is fully
developed once one has entered the plastic dynamical phase (as
defined from the critical analysis above) - appeals for a better
understanding of the latter. Characterization of the
instantaneous structure, such as the structure factor displayed
above, is useless to the study of the plastic phase: the
autocorrelation function of the instantaneous vortices positions ,
$\mathcal{C}(\mathbf{k})=<\rho(\mathbf{r})\:\rho(\mathbf{r+k})>_{\mathbf{r}}$,
where
$\rho(\mathbf{r})=\sum_{i=1}^{n_{V}}\:\delta(\mathbf{r_{i}})$
only confirms an evidence for a liquid order (Fig.\ref{autocor}).
Considering the existence of two distinct vortices populations
\cite{faleski96}: a rapidly moving ensemble of vortices along
quasi static channels and quasi pinned ones, one may also define
the velocity-weighted autocorrelation function,
$\mathcal{C}_{V}(\mathbf{k})=<\rho_{V}(\mathbf{r})\:\rho_{V}(\mathbf{r+k})>_{\mathbf{r}}$
where
$\rho_{V}(\mathbf{r})=\sum_{i=1}^{N_{V}}\:\delta(\mathbf{r_{i}})\:\dot{r}_{i}$.
This essentially measures the correlation amongst the most mobile
vortices. As can be seen in Fig.\ref{autocor}, the function
evolves from the one of a liquid to the one characteristic of
isolated flux channels (two peaks at $\mathbf{k} = (\pm
a_{0},0)$) as $j$ decreases. Both $\mathcal{C}(\mathbf{k})$ and
$\mathcal{C}_{V}(\mathbf{k})$ show that second neighbor
correlations are strongly damped in the transverse direction.
However, correlations between vortices may be found that are less
demanding than the ones uncovered by the transformations of the
instantaneous lattice. The autocorrelation function of the
channels, defined as :
$\mathcal{C}_{C}(\mathbf{k})=<\rho_{t}(\mathbf{r})\:\rho_{t}(\mathbf{r+k})>_{\mathbf{r}}$
where $\rho_{t}(\mathbf{r})=\int_{0}^{t}\rho_{V}(\mathbf{r})\:dt$
and $t$ is a time large enough so that the most mobile vortices
have moved by a distance larger than $a_{0}$, provides evidence -
close to the transition - for stronger transverse correlations
between such channels than the ones uncovered by
$\mathcal{C}(\mathbf{k})$ or $\mathcal{C}_{V}(\mathbf{k})$
(Fig.\ref{autocor}). This means that channels tend to correlate in
the direction transverse to the average flux flow.

\begin{figure}
\begin{center}
\epsfig{file=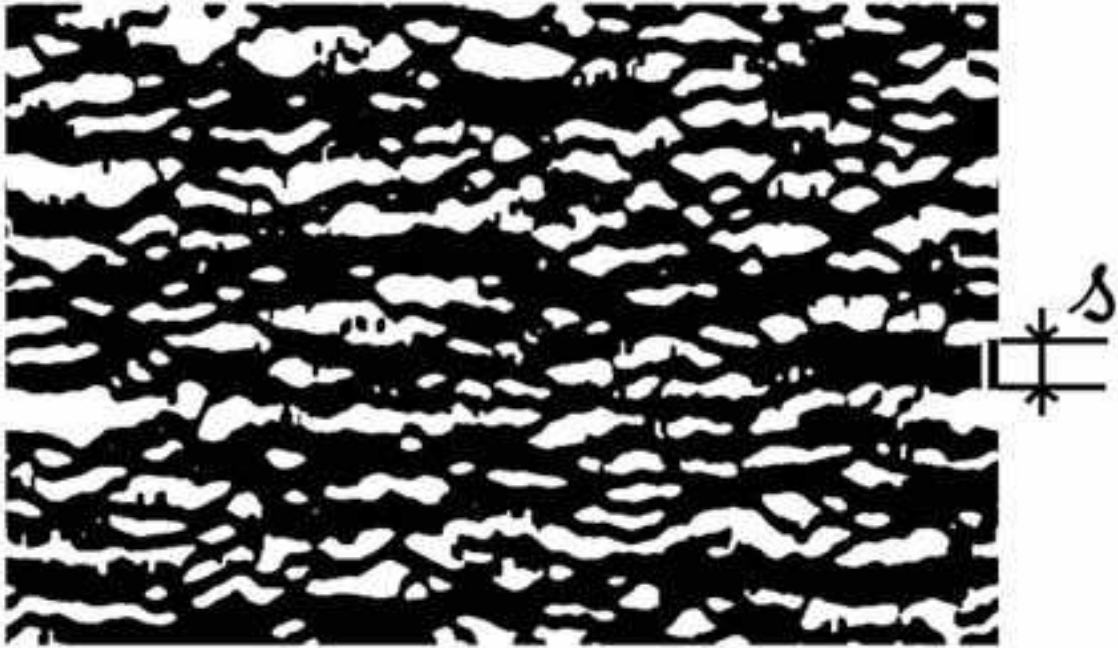, width=15cm}
\end{center}
\caption{Channels structure ($j=0.27$), filtered from frequencies
higher than $a_{0}^{-1}$. Shown as a white line is a
one-dimensional cluster of size \textit{s}.} \label{clusters}
\end{figure}

\begin{figure}
\begin{center}
\epsfig{file=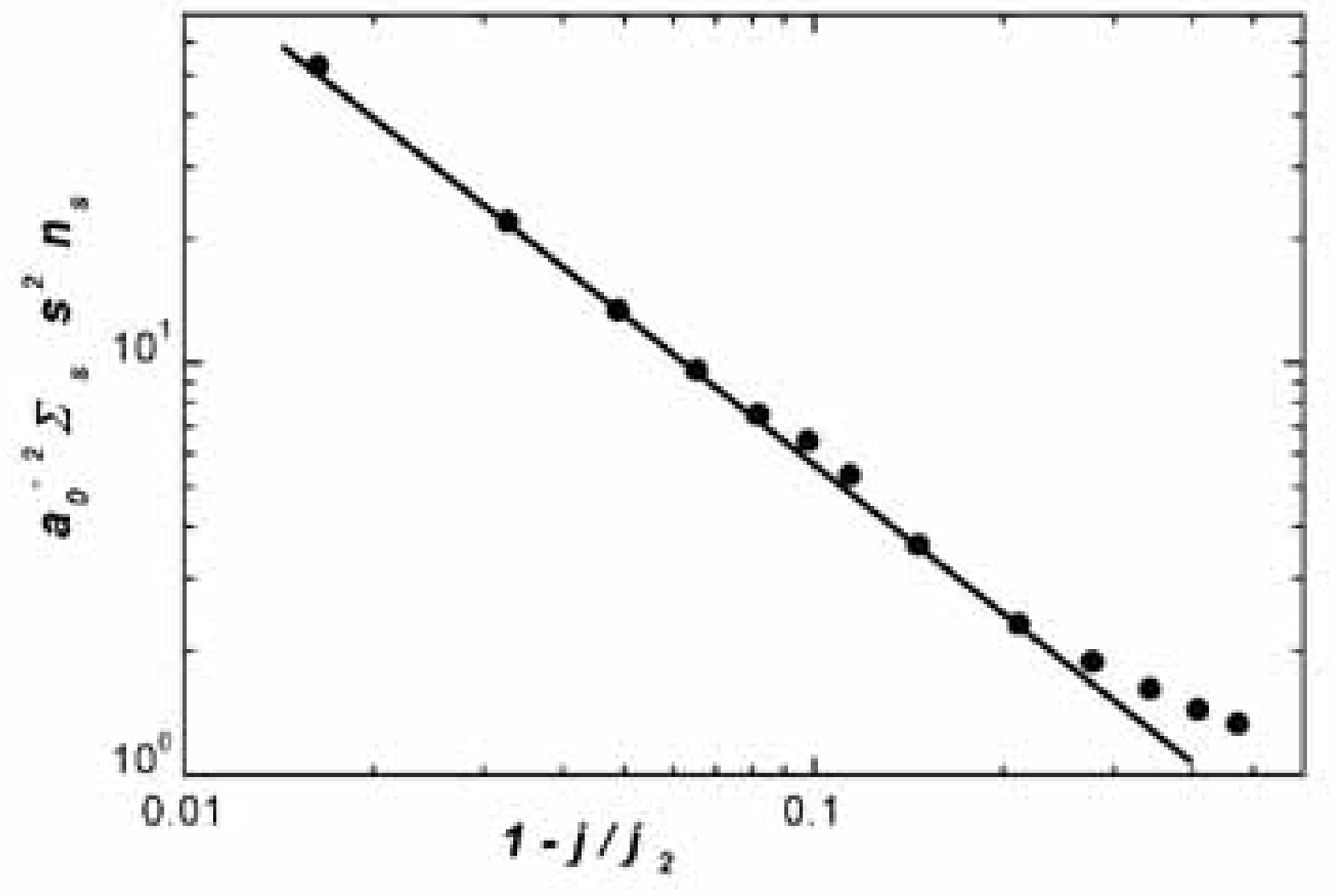, width=15cm}
\end{center}
\caption{Mean cluster size, normalized to $a_{0}^{2}$. The line is
a fit to $(1-j/j_{2})^{-\gamma}$ with $\gamma = 1.2 \pm 0.02$}
\label{scale}
\end{figure}

The qualitative analogy with percolation and the definition of a
critical order parameter, as shown above, both appeal for a
definition of dynamic clusters. This is done in the following
way: first, $\rho_{t}(\mathbf{r})$ is computed as defined above,
thus providing some snapshot of the channels. Then, the pattern
defined in this way is filtered from frequencies larger than
$a_{0}^{-1}$ (this insures that two contiguous channels do belong
to the same cluster). Finally, one-dimensional clusters are
defined, as the line segments perpendicular to the average flux
flow that are entirely contained within the filtered channels
(Fig.\ref{clusters}). Such a definition takes into account the
anisotropy of the problem and insures that an infinite cluster is
found at the transition. Although it provides an infinite number
of clusters for each sample, it allows the study of clusters
distributions as commonly done in the study of percolation
\cite{stauffer85}. The first infinite cluster is found at $j =
j_{2}$, in agreement with the critical analysis of the
resistivity. As shown in Fig.\ref{scale}, it is found that the
mean cluster size, $S=\sum_{s}s^{2}\;n_{s}$ scales close to
$j_{2}$ as $(1-j)^{-\gamma}$ with $\gamma = 1.2\pm0.02$. For $j <
0.2$, the mean cluster size saturates to $S = a_{0}^{2}$, meaning
that one enters a regime of isolated flux channels. This agrees
with $\mathcal{C}_{C}(\mathbf{k})$ in Fig.\ref{autocor}, where it
is seen that the first neighbor correlation roughly become
isotropic below $j \approx 0.2$. Following the analogy with
percolation, one may also define a correlation length,$\;\xi$,
which diverges at $j_{2}$. Then, the saturation observed for $S$
may be directly interpreted as the decrease of $\xi$ down to the
average flux line spacing, $a_{0}$. Within this description, it
could be appealing to interpret the existence of the anomalous
interface layer as a 'proximity effect'. However, the order
parameter - as defined above - should in this case continuously
\textit{increase} from zero in the ordered phase, to the value of
the bulk over a distance comparable to $\xi$, whereas it is
anomalously large in the pinned layer. Also, the thickness of the
anomalous layer should strongly depend upon $j$, which is not
observed in Fig.\ref{vitesse}. A more plausible interpretation
for the effect is in fact a topological one. Channels transverse
wandering - an alternative view for the clusters distribution and
the fractal topology of the plastic phase - is strongly
suppressed at the interface with the ordered phase. Besides the
occurrence of the ordered phase, the occurrence of a pinned
region at the interface provides another way to pin the
transverse excursions of the vortices as imposed by the proximity
of the crystal structure, hence the observed effect. However,
this piece of explanation does not provide any estimation for the
width of the layer. This shows that, although the present
analysis provides some evidence for the existence of a dynamical
plastic phase and a second order like transition, we still lack a
complete understanding for the pinned, driven vortex lattice.

I gratefully acknowledge valuable help from S. Ravy in the
handling of numerical diffraction data.

\begin{references}

\bibitem{fukuyama72} H. Fukuyama and P.A. Lee
Phys. Rev. B {\bf 17}, 535 (1972).

\bibitem{fisher85} D.S. Fisher, Phys. Rev. B {\bf 31}, 1396 (1985).

\bibitem{coppersmith91} S.N. Coppersmith and A.J. Millis,
Phys. Rev. B {\bf 44}, 7799 (1991).

\bibitem{koshelev94} A.E. Koshelev and V.M. Vinokur,
Phys. Rev. Lett. {\bf 73}, 3580 (1994).

\bibitem{brandt83} E.H. Brandt,
Phys. Rev. Lett. {\bf50}, 1599 (1983).

\bibitem{yaron95}U. Yaron, P.L. Gammel, D.A. Huse,
R.N. Kleiman, C.S. Oglesby, E. Bucher, B. Batlogg, D.J. Bishop,
K. Mortensen, K.N. Clausen, Nature. {\bf376}, 753 (1995).

\bibitem{moon96} K. Moon, R.T. Scalettar and G.T. Zimanyl,
Phys. rev. Lett. {\bf77}, 2778 (1996).

\bibitem{ledoussal98} P. LeDoussal and T. Giamarchi,
Phys. Rev. B {\bf 57}, 11356 (1998).

\bibitem{kolton99} A.B. Kolton, D. Dominguez and N.
Gronbech-Jensen, Phys. Rev. Lett. {\bf83}, 3061 (1999).

\bibitem{feigel89} M. Feigel'man, V. Geshkenbeim, A. Larkin and V.
Vinokur, Phys. Rev. Lett. {\bf 63}, 2303 (1989).

\bibitem{ryu96} R. Seungoh, M. Hellerqvist, S. Doniach, A.
Kapitulnik and D. Stroud, Phys. Rev. Lett. {\bf77}, 5114 (1996).

\bibitem{olson98} C.J. Olson,, C. Reichhardt and F. Nori,
Phys. Rev. Lett. {\bf81}, 3757 (1998).

\bibitem{brandt83a} E.H. Brandt, J. Low Temp. Phys. {\bf 53}, 41
(1983).

\bibitem{fangohr00} H. Fangohr, A.R. Rice, S.J. Cox, P.A.J. de
Groot, G.J. Daniell and K.S. Thomas, Journ. Comput. Phys.
{\bf162}, 372 (2000).

\bibitem{faleski96} M.C. Faleski, M.C. Marchetti and A.A. Middleton, Phys. Rev. B
{\bf54}, 12427 (1996).

\bibitem{stauffer85} D. Stauffer, 'Introduction to percolation
theory, Taylor and Francis (1985).


\end{references}
\end{document}